\documentclass[pre,preprint,showpacs,eqsecnum]{revtex4}
\usepackage{graphicx}
\begin{document}
\title{Analytically solvable model of a driven system with quenched
dichotomous disorder}
\author{S.~I.~Denisov,$^{1,2}$ M.~Kostur,$^{1}$ E.~S.~Denisova,$^{2}$
and P.~H\"{a}nggi$^{1,3}$} \affiliation{$^{1}$Institut f\"{u}r Physik,
Universit\"{a}t Augsburg, Universit\"{a}tsstra{\ss}e 1, D-86135 Augsburg,
Germany\\
$^{2}$Sumy State University, 2 Rimsky-Korsakov Street, 40007 Sumy, Ukraine\\
$^{3}$Department of Physics, National University of Singapore, Singapore
117542, Republic of Singapore}


\begin{abstract}
We perform a time-dependent study of  the driven dynamics of overdamped
particles which are placed in a one-dimensional, piecewise linear random
potential. This set-up of spatially quenched disorder then exerts a dichotomous
varying  random force on the particles. We derive the path integral
representation of the resulting  probability density function for the position
of the particles and transform this quantity of interest into the form of a
Fourier integral. In doing so, the  evolution of the probability density can be
investigated analytically for finite times. It is demonstrated that the
probability density contains both a $\delta$-singular contribution and a
regular part. While the former part plays a dominant role at short times, the
latter rules the behavior at large evolution times. The slow approach of the
probability density to a limiting Gaussian form as time tends to infinity is
elucidated in detail.
\end{abstract}

\maketitle

\section{INTRODUCTION}

The Langevin equation for an overdamped particle in a one-dimensional random
potential presents a useful model for the study of a variety of the
noise-induced and disorder-induced phenomena. Specifically, it has been
employed for describing anomalous diffusion in disordered media \cite{BG,BCGD},
glassy dynamics and depinning transition \cite{Sch,DV, Hor,GB, DH, LV},
diffusive transport in ratchets \cite{RH,AH,HMN} containing explicitly quenched
disorder \cite{Mar, PASF, GLZH}, carrier mobility in disordered solids
\cite{PKDK,KKDP}, to name only a few main ones. Moreover, this set-up enables a
very simple description of the dynamics of domain walls in random magnets,
dislocations in solids, and vortices in type-II superconductors, effectively
capturing all the essential effects of quenched disorder.

The effects arising from quenched disorder are often unexpected and even
counterintuitive. It therefore constitutes a prominent objective to study model
systems for which \textit{exact} statistical characteristics of the solution of
the corresponding Langevin equation can be obtained, not only at asymptotically
large times but even at finite times of its evolution. Exact results are very
useful also for testing approximate methods of analysis of these systems. There
already exists in the literature a number of exact results derived for systems
with Sinai disorder \cite{BG, BCGD, Sin, Gol, Mon}, Gaussian disorder
\cite{Sch, DV, Hor, GB,LV, HTB} and non-Gaussian disorder \cite{Den, DL, PKDK,
KKDP, DH}. In most of these cases the obtained results represent either the
\textit{numerical} characteristics of the solution of the Langevin equation
(moments of the solution, moments of the first-passage time, etc.) or the
\textit{asymptotic} probability distributions describing the long-time behavior
of the solution. To the best of our knowledge, however, the \textit{exact
time-dependent distributions} of analytic form, allowing the study of the
effects of quenched disorder in full detail, have not been given before.

In this paper we solve the above-mentioned challenge for a special class of
systems containing quenched \textit{dichotomous} disordered and subjected to a
constant bias force. More precisely, we derive and investigate the analytical
expression for the probability density function of overdamped particles in a
piecewise linear random potential driven by a constant force at zero
temperature. The paper is organized as follows. In Sec.~II, we describe the
model and introduce the path integral representation of the probability
density. In Sec.~III, we reduce the probability density to the form of a
Fourier integral. The first two moments of the probability density and their
asymptotic expressions are calculated in Sec.~IV. The time evolution of the
probability density is studied both analytically and numerically within Sec.~V.
Here we also compare the analytical results with those obtained from direct
numerical simulation and derive the asymptotic behavior of the probability
density in the limits of small and large times. In Sec.~VI, we summarize our
findings. Some technical details of our calculations are deferred to the
Appendixes A and B.

\section{BASIC EQUATIONS AND DEFINITIONS}

We consider the overdamped dynamics of a classical particle governed by the
dimensionless equation of motion
\begin{equation}
    \dot{X}_{t} = f + g(X_{t}).
    \label{eq motion}
\end{equation}
Here, $X_{t}$ denotes the particle coordinate, $f(>0)$ is a constant force, and
$g(x) =  -dU(x) / dx$ is the dichotomous random force produced by the piecewise
linear random potential $U(x)$ (see Fig.~1). We assume that $U(x)$ is
characterized by (i) statistically independent random intervals of lengths
$s_{j}$ which are distributed with the identical probability density $p(s)$ and
(ii) two deterministic potential slopes $+g$ and $-g$, so that $g(x) = \mp g$.
In addition, we assume that at the initial time $t=0$ a particle is located at
the origin of the coordinate system and that all sample paths of $U(x)$ start
with a positive slope $dU(x)/dx|_{x=+0} = +g$, so that
\begin{equation}
    X_{0} = 0, \quad \; g(+0) = -g.
    \label{init}
\end{equation}
Under these assumptions the dynamical solution of Eq. (\ref{eq
motion}) exists only if $f>g$; otherwise, if $f \leq g$, the
particle stays localized in its initial well.

Equation (\ref{eq motion}) is of minimal form in order to account for the
effects of quenched disorder on the overdamped motion of driven particles. As
will be demonstrated below, its main advantage is that many of the statistical
properties of $X_{t}$ can be derived exactly. Moreover, Eq.~(\ref{eq motion})
can be used also for studying a number of important physical issues ranging
from the low-temperature dynamics of charge carriers in randomly layered media
to the light propagation in interstellar space.

Our main objective is to express the probability density $P_{t}(x)$ that $X_{t}
= x$ for a fixed $t$ via the dimensionless characteristics of the force field.
In order to calculate this quantity, we start out from its definition
\begin{equation}
    P_{t}(x) = \langle \delta(x - X_{t}) \rangle,
    \label{def P}
\end{equation}
wherein the angular brackets denote an average over the sample paths of $g(x)$
and $\delta(x - X_{t})$ is the Dirac $\delta$-function. We first introduce the
total probability $W_{n}(t)$ of those sample paths which have $n(\geq 1)$ jumps
on the interval $(0,X_{t})$. In this case the particle coordinate $X_{t}$ lies
between the $n$th and $(n+1)$st jumps of $g(x)$ and thus can be represented as
\begin{equation}
    X_{t} = \sum_{j = 1}^{n}s_{j} + \tilde{s}_{n+1}
    \label{Xt}
\end{equation}
where $\tilde{s}_{n+1} < s_{n+1}$. Because, according to Eq.~(\ref{eq motion}),
the particle passes the interval $s_{j}$ during the time $s_{j}/[f +
(-1)^{j}g]$, the relation
\begin{equation}
    \sum_{j=1}^{n}\frac{s_{j}}{f + (-1)^{j}g} + \frac{\tilde{s}_{n+1}}
    {f + (-1)^{n+1}g} = t
    \label{rel1}
\end{equation}
holds. It implies that
\begin{equation}
    \tilde{s}_{n+1} = [f + (-1)^{n+1}g]\bigg(t - \sum_{j=1}^{n}
    \frac{s_{j}}{f + (-1)^{j}g} \bigg).
    \label{s n+1}
\end{equation}
The condition
\begin{equation}
    \sum_{j=1}^{n}\frac{s_{j}}{f + (-1)^{j}g} \leq t,
    \label{Omega n}
\end{equation}
being  a consequence of the relation $\tilde{s}_{n+1} \geq 0$, defines in the
$n$-dimensional space of the parameters $s_{j}$ a domain $\Omega_{n}(t)$ for
their allowed values. Therefore, since the probability that the $(n+1)$st jump
of $g(x)$ occurs at $s> \tilde{s}_{n + 1}$ is equal to $\int_{\tilde{s}_{n+1}}
^{\infty}p(s)ds$, we readily obtain
\begin{equation}
    W_{n}(t) = \int_{\Omega_{n}(t)}\bigg(\prod_{j=1}^{n}ds_{j}p(s_{j}) \bigg)
    \int_{\tilde{s}_{n+1}}^{\infty}p(s)ds.
    \label{Wn}
\end{equation}
If the function $g(x)$ has no jumps on the interval $(0,X_{t})$, i.e., if $n =
0$, then the total probability of these sample paths is given by
\begin{equation}
    W_{0}(t) = \int_{\tilde{s}_{1}}^{\infty}p(s)ds,
    \label{W0}
\end{equation}
where $\tilde{s}_{1} = (f-g)t$. We note that the probabilities $W_{0}(t)$ and
$W_{n}(t)$ are properly normalized: $W_{0}(t) + \sum_{n = 1}^{\infty}W_{n}(t) =
1$ (see Appendix A).

The above results, together with the definition (\ref{def P}), show that the
probability density function can be written  as
\begin{equation}
    P_{t}(x) = \delta[x - (f-g)t]W_{0}(t) + \sum_{n=1}^{\infty}
    P_{t}^{(n)}(x)\;,
    \label{P1}
\end{equation}
where
\begin{eqnarray}
    P_{t}^{(n)}(x) \!&=&\! \int_{\Omega_{n}(t)}\bigg(\prod_{j=1}^{n}
    ds_{j}p(s_{j}) \bigg) \int_{\tilde{s}_{n+1}}^
    {\infty}p(s)ds
    \nonumber\\[3pt]
    &&\! \times \delta(x - \sum_{j = 1}^{n}s_{j} - \tilde{s}_{n+1}).
    \label{def Pn}
\end{eqnarray}
These formulas provide a path integral representation of $P_{t}(x)$ in the case
of a dichotomous varying random force $g(x)$. We emphasize that this
representation is quite general and remains valid for an arbitrary probability
density $p(s)$ of the random intervals $s_{j}$.

\section{TIME-DEPENDENT PROBABILITY DENSITY FUNCTION}

From a practical point of view, the result (\ref{P1}) is inconvenient because
of its complexity. The main difficulty arises from the necessity to perform the
integration over the $n$-dimensional domain $\Omega_{n}(t)$ and the summation
over all $n$. Here we solve this problem in the case when the lengths of  the
random intervals $s_{j}$ are distributed with the exponential probability
density $p(s) = \lambda \exp(-\lambda s)$, where $\lambda ^{-1}$ denotes  the
average length of these intervals. The key point of our approach consists in
the use of the integral representation of the unit step function \cite{PBM}
\begin{equation}
    \frac{1}{2\pi}\int_{-\infty}^{+\infty}\frac{e^{(iz +y)\kappa}}
    {iz + y}\,dz = \left\{ \begin{array}{ll} 1 \;\; \textrm{if} \;
    \kappa > 0 \\ [6pt] 0 \;\;\textrm{if} \; \kappa < 0
    \end{array}
    \right.,
    \label{int1}
\end{equation}
where $y$ is a positive-valued, real parameter. Applying
(\ref{int1}) to (\ref{def Pn}) and setting  $\kappa = t -
\sum_{j=1}^{n} s_{j}/[f + (-1)^{j}g]$, we obtain
\begin{eqnarray}
    P_{t}^{(n)}(x) \!\!&=&\!\! \frac{1}{2\pi}\int_{-\infty}^{\infty}dz
    \frac{e^{(iz +y)t}}{iz + y}\int_{0}^{\infty}\!\ldots\!\int_{0}^{\infty}
    \!\bigg(\prod_{j=1}^{n} ds_{j} p(s_{j})
    \nonumber\\[6pt]
    && \times e^{-(iz +y)s_{j}/[f + (-1)^{j}g]}\bigg)\int_{\tilde{s}_{n +
    1}}^{\infty}p(s)ds
    \nonumber\\[3pt]
    && \times \delta(x - \sum_{j = 1}^{n}s_{j} - \tilde{s}_{n+1}).
    \label{Pn1}
\end{eqnarray}

Though in (\ref{Pn1}) an extra integral appears, this form of
$P_{t}^{(n)}(x)$ is much more advantageous as compared to the
integration over the domain $\Omega_{n}(t)$ in (\ref{P1}) because it
contains the independent integrations over the $n$ variables
$s_{j}$. Using the integral representation of the $\delta$-function
\begin{equation}
    \delta(\cdot) = \frac{1}{2\pi} \int_{-\infty}^{\infty}e^{-i\eta
    (\cdot)}d\eta
    \label{int2}
\end{equation}
and taking into account that according to (\ref{rel1})
\begin{equation}
    \sum_{j=1}^{2m}s_{j} + \tilde{s}_{2m+1} = (f - g)t + \frac{2g}{f+g}
    \sum_{j=1}^{m}s_{2j},
    \label{rel2}
\end{equation}
the formula (\ref{Pn1}) for $p(s) = \lambda \exp(-\lambda s)$ and $n
= 2m$ can be rewritten in the form
\begin{eqnarray}
    P_{t}^{(2m)}(x) \!&=&\! \frac{1}{(2\pi)^{2}}\int_{-\infty}^
    {\infty}d\eta\, e^{-i\eta[x - (f-g)t]}\int_{-\infty}^
    {\infty}\frac{dz}{iz + y}
    \nonumber\\[6pt]
    &&\!\times e^{[iz + y -\lambda(f-g)]t}I^{m}(\nu_{1})I^{m}(\nu_{2}).
    \label{P2m}
\end{eqnarray}
Here,
\begin{equation}
    I(\nu) = \int_{0}^{\infty} p(s)e^{-(\nu - \lambda)s}ds =
    \frac{\lambda}{\nu}
    \label{def I}
\end{equation}
with $\text{Re}\, \nu > 0$ and
\begin{equation}
    \nu_{1} = \frac{iz + y}{f - g}, \quad \nu_{2} = \frac{iz + y +
    2g(\lambda - i\eta)}{f + g}.
    \label{nu1,2}
\end{equation}
Similarly, since
\begin{equation}
    \sum_{j=1}^{2m-1}s_{j} + \tilde{s}_{2m+1} = (f + g)t - \frac{2g}{f-g}
    \sum_{j=1}^{m}s_{2j-1},
    \label{rel3}
\end{equation}
the formula (\ref{Pn1}) for $n = 2m-1$ yields
\begin{eqnarray}
    P_{t}^{(2m-1)}(x) \!&=&\! \frac{1}{(2\pi)^{2}}\int_{-\infty}^
    {\infty}d\eta\, e^{-i\eta[x - (f+g)t]}\int_{-\infty}^{\infty}
    \frac{dz}{iz + y}
    \nonumber\\[6pt]
    &&\!\times e^{[iz + y -\lambda(f+g)]t}I^{m}(\nu_{3})I^{m-1}(\nu_{4})
    \label{P2m-1}
\end{eqnarray}
with
\begin{equation}
    \nu_{3} = \frac{iz + y - 2g(\lambda - i\eta)}{f - g}, \quad
    \nu_{4} = \frac{iz + y}{f + g}.
    \label{nu3,4}
\end{equation}

Using the method of contour integration, it is easy to show that the integrals
over $z$ in (\ref{P2m}) and (\ref{P2m-1}) do not depend on the arbitrary
positive parameter $y$. This means that, in accordance with the definition
(\ref{def Pn}), the partial probability densities $P_{t}^{(2m)}(x)$ and
$P_{t}^{(2m - 1)}(x)$ do not depend on $y$ as well. Therefore, for calculating
the probability density $P_{t}(x)$ we can use the most appropriate values of
this parameter. Specifically, if before the integration over $z$ we want to sum
over $m$, it is reasonable to choose $y > \lambda (f + g)$ in order to avoid
dealing with divergent sums and integrals. In this case the condition
$\text{Re}\, \nu_{k} > \lambda$ holds for all $k$, and thus $|I(\nu_{k})| < 1$.
The last inequality permits one to use a geometric series formula for the
evaluation of series like
\begin{eqnarray}
    &\displaystyle\sum_{m=1}^{\infty}I^{m}(\nu_{1})I^{m}(\nu_{2}) =
    \frac{I(\nu_{1})I(\nu_{2})}{1 - I(\nu_{1})I(\nu_{2})},&
    \nonumber\\[6pt]
    &\displaystyle\sum_{m=1}^{\infty}I^{m}(\nu_{3})I^{m-1}(\nu_{4}) =
    \frac{I(\nu_{3})}{1 - I(\nu_{3})I(\nu_{4})}.&
    \label{uv2}
\end{eqnarray}
Then, substituting $W_{0}(t) = e^{-\lambda(f - g)t}$ and
\begin{equation}
    \sum_{n=1}^{\infty}P_{t}^{(n)}(x) = \sum_{m=1}^{\infty}
    [P_{t}^{(2m)}(x) + P_{t}^{(2m-1)}(x)]
    \label{ident1}
\end{equation}
into (\ref{P1}) and collecting the above derived results, we find the
probability density $P_{t}(x)$ in the form of the Fourier integral, i.e.,
\begin{equation}
    P_{t}(x) = \frac{1}{2\pi}\int_{-\infty}^{\infty}\phi_{t}(\eta)
    \,e^{-i\eta x}d\eta,
    \label{P2}
\end{equation}
where
\begin{eqnarray}
    \displaystyle \phi_{t}(\eta) \!\!&=&\!\! -\frac{e^{-(\lambda - i\eta)ft}}
    {2\pi}\int_{-\infty}^{\infty}\!\bigg[\frac{iz+y+2g(\lambda - i\eta)}
    {(z-z_{1})(z-z_{2})}\, e^{(\lambda - i\eta)gt}
    \nonumber\\[6pt]
    && \!\! + \frac{\lambda(f-g)}{(z-z_{3})(z-z_{4})}\,e^{-(\lambda -
    i\eta)gt}\bigg]e^{(iz + y)t}dz
    \label{G1}
\end{eqnarray}
is the characteristic function of $X_{t}$,
\begin{eqnarray}
    &z_{1} = g\eta + a(\eta) + i[y + \lambda g + b(\eta)],&
    \nonumber\\[3pt]
    &z_{2} = g\eta - a(\eta) + i[y + \lambda g - b(\eta)],&
    \nonumber\\[3pt]
    &z_{3} = -g\eta + a(\eta) + i[y - \lambda g + b(\eta)],&
    \nonumber\\[3pt]
    &z_{4} = -g\eta - a(\eta) + i[y - \lambda g - b(\eta)],&
    \label{z1-4}
\end{eqnarray}
and $a(\eta) = \lambda g^{2}\eta/b(\eta)$,
\begin{equation}
    b(\eta) = \frac{1}{\sqrt{2}}\{\lambda^{2}f^{2} - g^{2}\eta^{2}
    + [(\lambda^{2}f^{2} - g^{2}\eta^{2})^{2} + 4
    \lambda^{2}g^{4}\eta^{2}]^{\frac{1}{2}}\}^{\frac{1}{2}}.
    \label{def b}
\end{equation}

Finally, calculating the integral in (\ref{G1}) by the method of contour
integration (see Appendix B), we find the characteristic function
$\phi_{t}(\eta)$ in terms of elementary functions
\begin{eqnarray}
    \phi_{t}(\eta) \!&=&\! e^{-(\lambda - i\eta)ft}\bigg(
    \frac{\lambda f - ig\eta}{b(\eta)-ia(\eta)}\sinh[b(\eta)t-ia(\eta)t]
    \nonumber\\[3pt]
    && \!+ \cosh[b(\eta)t-ia(\eta)t]\bigg).
    \label{G2}
\end{eqnarray}
As expected, the characteristic function does not depend on the auxiliary
parameter $y$. The functions $a(\eta)$ and $b(\eta)$ tend to $g\eta$ and
$\lambda g$, respectively, as $|\eta| \to \infty$. Accordingly, $\phi_{t}(\eta)
\propto e^{i \eta (f-g)t}$ if $|\eta| \to \infty$, and so the probability
density $P_{t}(x)$ contains a $\delta$-singular part. This fact can be
displayed in the explicit way as follows:
\begin{equation}
    P_{t}(x) = e^{-\lambda(f-g)t}\delta[x - (f-g)t] +
    \tilde{P}_{t}(x),
    \label{P3}
\end{equation}
where
\begin{equation}
    \tilde{P}_{t}(x) = \frac{1}{2\pi}\int_{-\infty}^{\infty}
    \tilde{\phi}_{t}(\eta)\,e^{-i\eta x}d\eta
    \label{tilde P}
\end{equation}
and
\begin{equation}
    \tilde{\phi}_{t}(\eta) = \phi_{t}(\eta) - e^{-(f-g)(\lambda -
    i\eta)t} \;
    \label{tilde G}
\end{equation}
with $\tilde{\phi}_{t}(\eta) \to 0$ as $|\eta| \to \infty$. The formulas
(\ref{G2})-(\ref{tilde G}), which give the probability density $P_{t}(x)$ in a
much simpler  and transparent form as compared to (\ref{P1}) and (\ref{def
Pn}), constitute our main results. To the best of our knowledge, this is the
first example of nontrivial system with quenched disorder for which the
statistical properties of the particle coordinate $X_{t}$ can be studied in
full detail.

The above derived findings can be used also for studying the probability
density in the case of the dichotomous random function $g(x)$ possessing
statistical properties different from the especially chosen initial condition.
In particular, if $g(+0) = -g$ with a probability $p_{1}$ and $g(+0) = +g$ with
a probability $p_{2}$ which satisfy the condition $p_{1} + p_{2} =1$, then the
corresponding probability density $R_{t}(x)$ is given by $R_{t}(x) =
p_{1}P_{t}(x) + p_{2}P_{t}(x) |_{g \to -g}$.

\section{MOMENTS OF THE PROBABILITY DENSITY}

The moments of the probability density, $\langle X_{t}^{k} \rangle =
\int_{-\infty}^{\infty}x^{k} P_{t}(x) dx$ $(k = 1,2,\ldots)$, are expressed
through the characteristic function in the commonly known way:
\begin{equation}
    \langle X_{t}^{k} \rangle = \frac{1}{i^{k}} \frac{d^{k}}
    {d\eta^{k}}\,\phi_{t}(\eta)\Big|_{\eta = 0}.
    \label{<X^k>}
\end{equation}
In principle, with the help of Eqs.~(\ref{<X^k>}) and (\ref{G2}) any moment can
be calculated straightforwardly. Specifically, for the first moment these
equations yield
\begin{equation}
    \langle X_{t} \rangle = \frac{f^{2} - g^{2}}{f}\,t -
    \frac{g(f-g)}{2\lambda f^{2}}(1 - e^{-2\lambda ft}).
    \label{<X>}
\end{equation}
In the case of small times, obeying $t \ll 1/\lambda f$, this expression
reduces to $\langle X_{t} \rangle = (f-g)t$. It has a simple interpretation: At
small times the condition $g(X_{t}) = -g$ holds with almost unit probability
and so almost with probability one a particle moves with the dimensionless
velocity $f - g$. If $t \gg g/\lambda f^2$, then $\langle X_{t} \rangle =
(f^{2} - g^{2})t/f$, i.e., the long-time average velocity of a particle,
$\langle v \rangle = \lim_{t \to \infty} \langle X_{t} \rangle / t$, equals
$(f^2 - g^2)/f$ (see Fig.~2, a). In order to elucidate this result further, we
first note that the total average length of the odd intervals $s_{2m-1}$ (and
the even intervals $s_{2m}$) on the interval $(0,\langle X_{t} \rangle)$ tends
to $\langle X_{t} \rangle/2$ as $t \to \infty$. Accordingly, the total time
which a particle spends in the odd and even intervals are $t_{-} = \langle
X_{t} \rangle/ 2(f - g)$ and $t_{+} = \langle X_{t} \rangle/2(f + g)$,
respectively. Therefore, using the condition $t_{-} + t_{+} = t$, one obtains
in the long-time limit that $\langle v \rangle = (f^2 - g^2)/f$. We emphasize
that this result does not depend on the explicit form of the probability
density $p(s)$ of the random intervals $s_{j}$.

According to Eqs.~(\ref{<X^k>}) and (\ref{G2}), the second moment is given by
\begin{eqnarray}
    \langle X_{t}^{2} \rangle \!&=&\! \frac{(f^{2} - g^{2})^{2}}{f^{2}}
    \,t^{2} - \frac{g(f^2-g^2)(f-2g)}{\lambda f^{3}}\,t
    \nonumber\\[3pt]
    &&\! - \frac{g^2(f-g)(f+3g)}{2\lambda^2 f^4} +
    \frac{g(f-g)}{2\lambda^2 f^4}
    \nonumber\\[3pt]
    &&\! \times [2\lambda f(f^2+g^2)\,t + g(f+3g)]e^{-2\lambda ft}.\quad
    \label{<X^2>}
\end{eqnarray}
As a consequence, the variance $\sigma_{t}^{2} = \langle X_{t}^{2} \rangle -
\langle X_{t} \rangle ^{2}$ of the particle coordinate $X_t$ can be represented
in the form
\begin{eqnarray}
    \sigma_{t}^{2} \!\!&=&\!\! \frac{g^{2}(f - g)}{4\lambda^2 f^4}
    [4\lambda f(f+g)\,t - 3f - 5g + 4(f+g
    \nonumber\\[3pt]
    &&\!\! + 2\lambda fg\,t)\,e^{-2\lambda ft} - (f-g)\,e^{-4\lambda ft}].
    \label{sigma}
\end{eqnarray}
At small times, when $t \ll 1/\lambda f$, this position variance $\sigma_{t}
^{2}$ grows as $t^3$, $\sigma_{t}^{2} = (4/3)\lambda g^{2}(f-g)t^{3}$, and at
large times, when $t \gg 1/\lambda f$, it grows as $t$, $\sigma_ {t}^{2} =
g^{2}(f^2- g^2)\, t/\lambda f^3$. The last result evidences that the particles
exhibit normal biased diffusion with an effective diffusion coefficient
$D_{\text{eff}}$, reading:
\begin{equation}
    D_{\text{eff}} = \lim_{t \to \infty}\frac{\sigma_{t}^{2} }{2t} =
    \frac{g^{2}(f^{2} - g^{2})}{2\lambda f^{3}}.
    \label{Deff}
\end{equation}
Interestingly, $D_{\text{eff}}$ varies non-monotonically with the external
force $f$: $D_{\text{eff}} = (f - g)/ \lambda$ for $f \approx g$,
$D_{\text{eff}} = g^{2}/ 2\lambda f$ for $f \gg g$, and $\max {D_{\text{eff}}}
= g/ (3^{3/2} \lambda)$ for $f = \sqrt{3}\, g$ (see Fig.~2, b).

\section{TIME EVOLUTION OF THE PROBABILITY DENSITY}

According to Eqs.~(\ref{tilde G}) and (\ref{G2}), at $t = 0$ the relation
$\tilde {\phi}_{0}(\eta) = 0$ holds and thus $\tilde{P}_{0}(x) = 0$ and
$P_{0}(x) = \delta(x)$. This result is a direct consequence of the initial
condition $X_{0} = 0$. At $t> 0$ the probability density $P_{t}(x)$ contains
both a $\delta$-singular part and a regular part. The $\delta$-singular part
arises from the existence of a finite, nonzero probability that the random
function $g(x)$ does not change at all the sign on the interval $[0,(f-g)t]$.
The total probability of these sample paths, $W_{0}(t) = e^{-\lambda(f - g)t}$,
defines the weight of the $\delta$-singular distribution part and, because the
particles move with the velocity $f-g$, this distribution moves towards the
right with the same velocity.

Because the minimal and maximal velocities of the particles are $f-g$ and
$f+g$, respectively, the regular part of the probability density, $\tilde{P}
_{t}(x)$, is concentrated on the interval $[(f-g)t, (f + g)t]$. If $x$ belongs
to this interval, then $\lim_{t \to 0}P_{t}(x) = \lambda(f - g)/2g$. For small
but finite times $\tilde{P}_{t}(x)$ is an almost linear, decreasing function of
$x$ which is transformed to a unimodal probability density as $t$ increases.
These features of the probability density are illustrated in Fig.~3. In order
to verify and test the theory for calculating $P_{t}(x)$, we performed as well
direct numerical simulation of Eq.~(\ref{eq motion}). As depicted in Fig.~4,
our analytical results are in excellent agreement with the numerical findings.

\subsection{Asymptotic Approach to a Gaussian Density}

Based on the central limit theorem of probability theory, see, e.g., Ref.
\cite{GK}, we may expect that the probability density $P_{t} (x)$ tends to the
normal (Gaussian) probability density as $t \to \infty$. To gain more insight
into the long-time behavior of $P_{t}(x)$, we consider the scaled probability
density defined  by
\begin{equation}
\mathcal{P}_{t}(\xi) = \sigma_{t} P_{t}(\langle X_{t} \rangle +
\sigma_{t} \xi)
 \label{PSCALED}
\end{equation}
which, in accordance with (\ref{P2}), can be written in the form
\begin{equation}
    \mathcal{P}_{t}(\xi) = \frac{1}{2\pi} \int_{-\infty}^{\infty}
    \phi_{t}( u/\sigma_{t})\,e^{-iu\langle X_{t} \rangle
    /\sigma_{t} - iu\xi}du.
    \label{P4}
\end{equation}
Assuming that $|u| \ll \sigma_{t}^{1/3}$ and keeping only the first two terms
of the asymptotic expansion of the function $\phi_{t}( u/\sigma_{t})\, e^{-iu
\langle X_{t} \rangle/\sigma_{t}}$ as $t \to \infty$, we obtain
\begin{equation}
    \phi_{t}( u/\sigma_{t})\,e^{-iu\langle X_{t} \rangle/\sigma_{t}}
    = e^{-u^{2}/2}\bigg[1 - i\frac{gu^{3}}{2\sqrt{\lambda f(f^{2} -
    g^{2})t}} \bigg].
    \label{G3}
\end{equation}
Substituting this formula into (\ref{P4}) yields $\mathcal{P}_{t}(\xi) =
\mathcal{P}_{\infty}(\xi) + \mathcal{P}_{t}^{(1)}(\xi)$, where
\begin{equation}
    \mathcal{P}_{\infty}(\xi) = \frac{1}{2\pi} \int_{-\infty}^{\infty}
    e^{-u^{2}/2 - iu\xi}du = \frac{e^{-\xi^{2}/2}}{\sqrt{2\pi}}
    \label{P norm}
\end{equation}
is the probability density of the standard normal distribution and
\begin{eqnarray}
    \mathcal{P}_{t}^{(1)}(\xi) \!\!&=&\!\! -i \frac{g^{2}}{4\pi
    \lambda f^{2}}\int_{-\infty}^{\infty} e^{-u^{2}/2 - iu\xi}u^{3}du
    \nonumber\\[3pt]
    \!\!&=&\!\! -\frac{g(3\xi - \xi^{3})}{2\sqrt{\lambda
    f(f^{2} - g^{2})t}} \,\mathcal{P}_{\infty}(\xi)
    \label{P(1)}
\end{eqnarray}
($|\mathcal{P} _{t}^{(1)}(\xi)|/ \mathcal{P}_{\infty}(\xi) \ll 1$)
describes the deviation of $\mathcal{P}_{t}(\xi)$ from $\mathcal{P}
_{\infty}(\xi)$. According to this result, $\mathcal{P}_{t}(\xi)$
slowly approaches ($\propto t^{-1/2}$)  the asymptotic Gaussian form
as $t \to \infty$.

A quantitative measure of non-Gaussian behavior of $P_{t} (x)$ can be
characterized by the kurtosis defined as
\begin{equation}
    k(t) = \frac{\langle(X_{t} - \langle X_{t} \rangle)^{4}\rangle}
    {\sigma_{t}^{4}} - 3.
    \label{k(t)}
\end{equation}
Notably, $k(t)$ is equal to zero if $X_{t}$ has the Gaussian distribution. Its
dependence on time is calculated with the help of Eq.~(\ref{<X^k>}) and is
depicted in Fig.~5. In accordance with the central limit theorem, the kurtosis
tends to zero as $t$ increases. At very small times the formula (\ref{P3})
yields
\begin{equation}
    P_{t}(x) = [1 - \lambda (f-g)t]\,\delta [x - (f - g)t] +
    \frac{\lambda (f - g)}{2g},\;
    \label{P5}
\end{equation}
where $x \in [(f-g)t,(f+g)t]$. For this probability density we readily obtain
$\langle(X_{t} - \langle X_{t} \rangle)^{4}\rangle = (16/5) \lambda g^{4} (f -
g)t^{5}$, $\sigma_{t}^{2} = (4/3)\lambda g^{2} (f - g)t^{3}$, and so $k(t) =
(9/5)[\lambda (f-g)t]^{-1}$ as $t \to 0$. The divergence of kurtosis as $t \to
0$ corroborates the fact that at small times the probability density indeed
strongly differs from the normal one.

The fact that $P_{t}(x)$ approaches asymptotically the normal probability
density function with $\sigma_{t} ^{2} \propto t$ as $t \to \infty$ leads to an
interesting conclusion: Under certain conditions the long-time behavior of the
considered particles is the same as the long-time behavior of Brownian
particles. In order to elucidate this, we use the dimensionless equation of
motion for overdamped Brownian particles: $\dot{x}_{t} = F + \xi(t)$ ($x_{0} =
0$). Here, $x_{t}$ is the Brownian particle coordinate, $F$ is an external
force, and $\xi(t)$ is Gaussian, thermal white noise with zero mean and
correlation function $\overline{\xi(t) \xi(t')} = 2\Delta \delta(t - t')$ (the
overbar denotes an average over all realizations of $\xi(t)$ and $\Delta$ is
the noise intensity). As is well-known, see, e.g., Refs. \cite{H-T,V-K},
$x_{t}$ has a Gaussian density with $\overline{x}_{t} = Ft$ and
$\overline{(x_{t} - \overline {x}_{t})^2} = 2\Delta t$. Therefore, the
long-time statistical properties of usual Brownian particles and particles in
our case are the same if $\overline{x}_{t} = \langle X_{t} \rangle$ and
$\overline{(x_{t} - \overline{x}_{t})^2} = \sigma_{t}^{2}$, i.e., $F = (f^{2} -
g^{2})/f$ and $\Delta = D_{\text{eff}}$. We further note that in the Brownian
case the internal noise intensity $\Delta$ is proportional to the absolute
temperature; thus, the long-time behavior of $X_{t}$ can also be characterized
by the dimensionless effective temperature $T_{\text{eff}} = D_{\text{eff}}$.

\subsection{Numerical Simulations}

The numerical simulation of Eq.~(\ref{eq motion}) turned out to be a valuable
tool in verifying our analytical findings. We take advantage of the fact that
this equation contains no explicit time dependent terms. In such a case, the
numerical simulations can be made especially efficient. According to
Eq.~(\ref{eq motion}), inside each interval where $g(x) = -1$ or $+1$ the
particle velocity is given by $f - g$ or $f + g$, respectively. Thus, the
algorithm consists of successive generations of random interval lengths
according to the exponential distribution and a calculation of the times needed
to travel the generated interval with constant velocity $f - g$ or $f + g$. The
total time is summed up until it reaches the required final value. The ensemble
average is then obtained by repeating this outlined  procedure with different
realizations of random intervals. The probability density function is then
obtained as the histogram of final positions of the particle. Using this
method, we could obtain the probability density $P_{t}(x)$ with similar
computational effort as the calculation of the Fourier integral (\ref{P2}). The
averaging over $10^7$ realizations usually assumed several seconds on modern
workstations. The existence of the $\delta$-singular part of $P_{t}(x)$ was
recognized in our histogram-procedure by changing the bin size of the
histogram: The number of counts in the histogram bin containing the
$\delta$-function does not depend on the bin size.

\section{DISCUSSION AND CONCLUSIONS}

We used a path integral approach for determining the probability density
function of overdamped particles in a piecewise linear random potential which
are in addition driven by a constant force. Assuming that the intervals of the
piecewise linear parts of the potential are distributed with an exponential
distribution, we succeeded in obtaining the time-dependent probability density
in the form of an explicit Fourier integral. Within this framework, we showed
that the probability density contains, apart from a regular part, also  a
$\delta$-singular contribution. The weight of this $\delta$-singular part
exponentially decreases with time and the total probability density slowly, as
$t^{-1/2}$, converges to a Gaussian density in the long-time limit.

We further calculated the first and the second moments of the probability
density function and studied its time evolution, both analytically and
numerically. Our analytical results are in perfect agreement with the numerical
ones obtained from simulations of the equation of motion. Moreover, we derived
a simple representation of the probability density for small and large times
and, to characterize the non-Gaussianity of the probability density, we
calculated the kurtosis as a function of time. We showed that under certain
conditions the long-time behavior of the considered particles is the same as
Brownian particles. The corresponding effective diffusion coefficient and
effective temperature are calculated.

\section*{ACKNOWLEDGMENTS}

S.I.D. acknowledges the support of the EU through contract No
MIF1-CT-2006-021533 and P.H. has been supported by the Deutsche
For\-schungs\-ge\-mein\-schaft via the Collaborative Research Centre SFB-486,
project A10.  Financial support of the German Excellence Initiative via the
{\it Nanosystems Initiative Munich} (NIM) is gratefully acknowledged as well.

\appendix

\section{Proof of the normalization condition for
$\textit{W}_{\textit{n}}\textbf{(}\textit{t}\textbf{)}$}

Let us introduce the quantities
\begin{equation}
    S_{n}(t) = \int_{\Omega_{n}(t)}\prod_{j=1}^{n}p(s_{j})ds_{j}
    \label{Sn1}
\end{equation}
with $n \geq 1$.  According to the definitions of $\Omega _{n}(t)$ and
$\tilde{s}_{k}$,  these quantities can be written in the form
\begin{equation}
    S_{n}(t) = \int_{0}^{\tilde{s}_{1}}p(s_{1})ds_{1}
    \int_{0}^{\tilde{s}_{2}}p(s_{2})ds_{2}\ldots
    \int_{0}^{\tilde{s}_{n}}p(s_{n})ds_{n}.
    \label{Sn2}
\end{equation}
Using in Eq.~(\ref{Wn}) the integral relation $\int_{\tilde{s}_{n+1}} ^{\infty}
p(s)ds = 1 - \int_{0}^{\tilde{s}_{n+1}}p(s_{n+1})ds_{n+1}$ resulting from the
normalization condition for $p(s)$, we can express the probabilities $W_{n}(t)$
through the quantities $S_{n}(t)$ as follows:
\begin{equation}
    W_{n}(t) = S_{n}(t) - S_{n+1}(t).
    \label{rep Wn}
\end{equation}
Taking also into account that $S_{\infty}(t) = 0$, this
representation of $W_{n}(t)$ yields $\sum_{n=1} ^{\infty} W_{n}(t) =
S_{1}(t)$. On the other hand, since $W_{0}(t) = 1 - S_{1}(t)$, we
find that the normalization condition $W_{0}(t) + \sum_{n=1}
^{\infty}W_{n}(t) = 1$ holds true.

\section{Derivation of Eq.~(\ref{G2})}

According to (\ref{G1}), the characteristic function $\phi_{t}(\eta)$ can be
written in the form
\begin{equation}
    \phi_{t}(\eta) = -e^{-(\lambda - i\eta)ft}
    [A(\eta)e^{(\lambda - i\eta)gt} + B(\eta)e^{-(\lambda - i\eta)gt}]
    \label{G4}
\end{equation}
with
\begin{equation}
    \displaystyle A(\eta) = \frac{1}{2\pi}\int_{-\infty}^{\infty}
    \frac{iz+y+2g(\lambda - i\eta)}{(z-z_{1})(z-z_{2})}\, e^{(iz + y)t}dz
    \label{Y1}
\end{equation}
and
\begin{equation}
    \displaystyle B(\eta) = \frac{1}{2\pi}\int_{-\infty}^{\infty}
    \frac{\lambda(f - g)}{(z-z_{3})(z-z_{4})}\, e^{(iz + y)t}dz.
    \label{Y2}
\end{equation}
Applying the residue theorem for calculating the integrals in (\ref{Y1}) and
(\ref{Y2}), we obtain
\begin{eqnarray}
    &A(\eta) = i[\text{Res}\,\Psi(\eta;z_{1}) + \text{Res}\,
    \Psi(\eta;z_{2})]\;&
    \nonumber\\[6pt]
    &B(\eta) = i[\text{Res}\,\Phi(\eta;z_{3}) + \text{Res}\,
    \Phi(\eta;z_{4})]\;.&
    \label{Y1,Y2}
\end{eqnarray}
Here, $\text{Res}\,\Psi(\eta;z_{1,2})$ and $\text{Res}\,\Phi(\eta;z_{3,4})$ are
the residues of the functions
\begin{eqnarray}
    &\displaystyle \Psi(\eta;z) = \frac{iz+y+2g(\lambda - i\eta)}{(z-z_{1})
    (z-z_{2})}\, e^{(iz + y)t},&
    \nonumber\\[6pt]
    &\displaystyle \Phi(\eta;z) = \frac{\lambda(f - g)}{(z-z_{3})(z-z_{4})}\,
    e^{(iz + y)t}&
    \label{PsiPhi}
\end{eqnarray}
of the complex variable $z$ at the points $z_{k}$. Assuming that $y> \lambda (f
+ g)$ (in this case $\text{Im}\,z_{k} > 0$) and taking into account that
$\text{Res} \,\Psi(\eta;z_{1,2}) = \lim_{z \to z_{1,2}}(z - z_{1,2})
\Psi(\eta;z)$ and $\text{Res}\,\Phi(\eta;z_{3,4}) = \lim_{z \to z_{3,4}}(z -
z_{3,4}) \Phi(\eta;z)$, from Eqs. (\ref{Y1,Y2}), (\ref{PsiPhi}) and
(\ref{z1-4}) we find
\begin{eqnarray}
    A(\eta) \!\!&=&\!\! -e^{-(\lambda - i\eta)gt}\bigg(
    \frac{g(\lambda -i\eta)}{b(\eta) - ia(\eta)}
    \sinh{[b(\eta)t - ia(\eta)t]}
    \nonumber\\[6pt]
    &&\!\! + \cosh{[b(\eta)t - ia(\eta)t]}\bigg),
    \nonumber\\[8pt]
    B(\eta) \!\!&=&\!\! \frac{\lambda (g - f)}{b(\eta) - ia(\eta)}
    e^{(\lambda - i\eta)gt}\sinh{[b(\eta)t - ia(\eta)t]}.\quad\quad
    \label{AB}
\end{eqnarray}
Finally, substituting (\ref{AB}) into (\ref{G4}), we end up with (\ref{G2}).

\newpage

\begin{figure}
    \centering
    \includegraphics{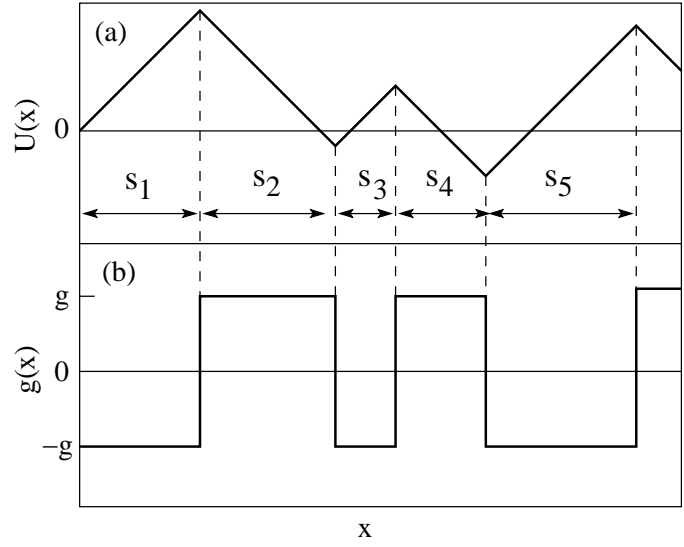}
    \caption{\label{fig1} Model potential and force field. (a) Schematic
    representation of the piecewise varying linear random potential
    $U(x)$  and  (b) the corresponding dichotomous random force
    $g(x) = -dU(x)/ dx$  as functions of the particle coordinate $x$. }
\end{figure}

\begin{figure}
    \centering
    \includegraphics{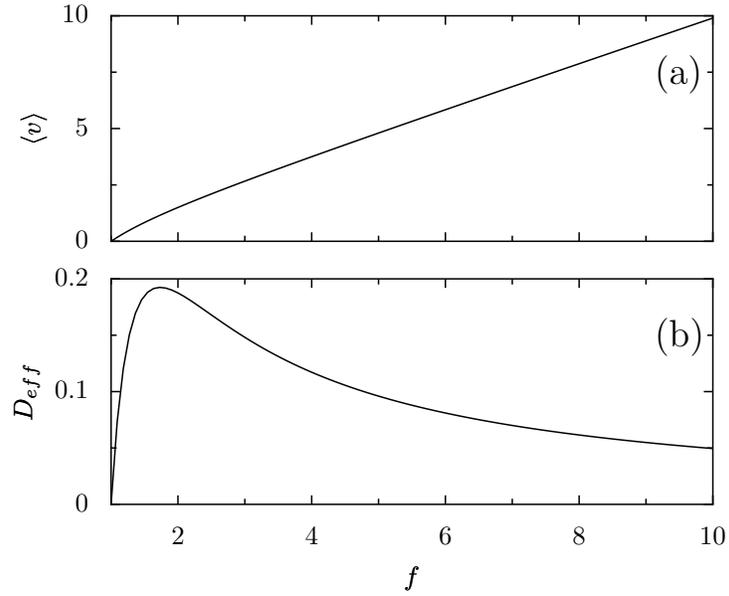}
    \caption{\label{fig2} (a) Behavior of the average velocity
    $\langle v \rangle$ and (b) of the effective diffusion coefficient
    $D_{\text{eff}}$ versus the external force $f$ for the chosen parameter
    values $\lambda = g = 1$.}
\end{figure}

\begin{figure}
    \centering
    \includegraphics{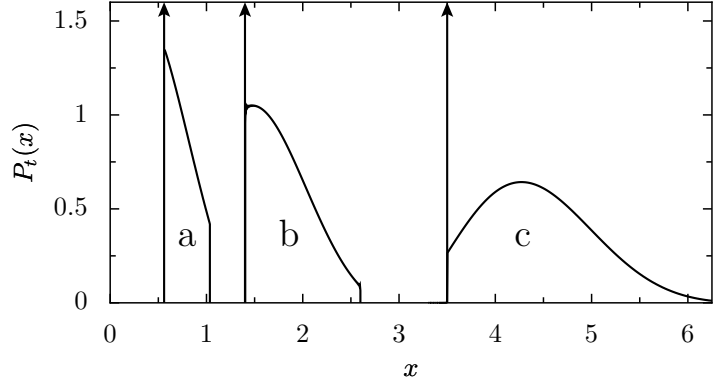}
    \caption{\label{fig3} Time evolution of the probability density.
    Depicted are the plots of the probability density $P_{t}(x)$ derived
    from Eqs.~(\ref{G2})-(\ref{tilde G}) for the evolution time (a)
    $t=0.75$, (b) $t=2$,  and (c) $t=5$. The values of the other
    parameters are set at $f=1$, $g=0.3$, and $\lambda=1$. The vertical
    arrows depict the singular parts of decreasing weight of $P_{t}(x)$.}
\end{figure}

\begin{figure}
    \centering
    \includegraphics{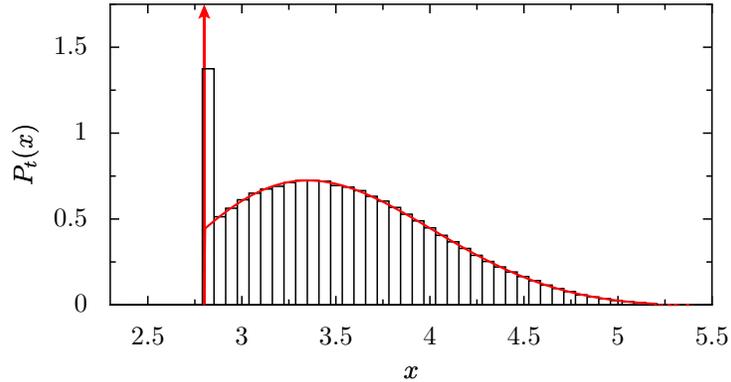}
    \caption{\label{fig4} Theory versus numerics. We show the probability
    density  $P_{t}(x)$ as a function of the coordinate $x$ for $f=1$, $g=0.3$,
    $\lambda=1$, and time $t=4$. The solid line (red online) represents
    the theoretical result obtained from Eqs.~(\ref{G2})-(\ref{tilde G}),
    while the histogram refers to our numerical simulation of
    Eq.~(\ref{eq motion}).}
\end{figure}

\begin{figure}
    \centering
    \includegraphics{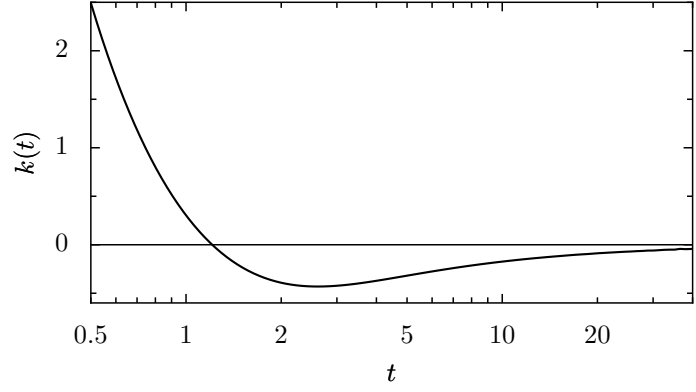}
    \caption{\label{fig5} Deviation from Gaussian behavior. The kurtosis
    $k(t)$ of the probability density is depicted on a linear [$k(t)$]
    versus logarithmic [$t$] scale as a function of the evolution time
    $t$ for $f=1$, $g=0.3$, and $\lambda=1$. }
\end{figure}

\end{document}